\documentstyle[pra,aps,graphicx]{revtex}
\begin{document}
\draft
\twocolumn[\hsize\textwidth\columnwidth\hsize\csname 
@twocolumnfalse\endcsname
\title{Transmission of Quantum Images Through Long Distances} 
\author{M. P. Almeida  and P.H. Souto Ribeiro$^{*}$}
\address{Instituto de F\'{\i}sica, Universidade Federal do Rio de 
Janeiro, Caixa Postal 68528, Rio de Janeiro, RJ 22945-970, Brazil}
\date{\today}
\maketitle
\begin{abstract}
We report on an experiment demonstrating the principle for transmitting
quantum images through long distances. Signal and idler beams carrying
correlated images have natural divergences that can be compensated by
the use of collimating lenses and at the same time preserving the information
contained in their correlated angular spectra.
\end{abstract}
]
\section{Introduction}
The use of the quantum entanglement for practical applications consists
in a formidable challenge. The twin photons produced in the parametric
down-conversion, can be prepared in entangled states that can be used
for instance, in quantum communication\cite{1}. Most of the projects related to the
use of entangled photons for quantum communication, rely on the manipulation
of the photon polarization as the entangled degree of freedom\cite{2}. In
particular, the capability of reliably transmitting an entangled photon pair through
a long distance have been pursued, with\cite{3} and without\cite{4} the use of
optical fibers.
Optical fibers are a very convenient and robust medium for transmitting optical signals. 
However, sometimes it is not possible to use it. One nice example is the project
for transmitting entangled photon pairs via satellite\cite{5}.  As a first approach,
it has been demonstrated that it is possible to share an entangled photon pair
through about 600m of free air\cite{6}. In all the cases mentioned, the entangled
degree of freedom was the polarization. In this paper, we are going to show that
it is also possible to share entangled photon pairs through long distances, manipulating
other degree of freedom: the transverse momentum of the photon.
Twin photons entangled in the transverse momentum give rise to the so called
{\it Quantum Images}. The term quantum image has been used for the correlated
images in optical fields\cite{7}. A discussion on the issue
of the quantum character of these correlated images is actually in progress\cite{8}.
In this work, we are going to deal with correlated images transferred from the
pump to the correlated angular spectrum\cite{9}. 
In this case, it is possible to show that some correlated images, have transverse correlations that violate a classical inequality, leading
to spatial photon anti-bunching\cite{10,11}. The present work is based in this 
kind of correlated images and we going to call them quantum images, even though
there may have place for further discussion on the issue.
We are going to demonstrate that it is possible to transmit a quantum image
through the free air for long distances. The scheme consists in compensating
the divergence of the signal and idler beams without loosing the spatial correlations
between them. First, we show how the use of collimating lenses increase the
coincidence counting rate spatial density and second we demonstrate the transmission
of a correlated image through about 3m.
\section{Theoretical Background}

\begin{figure}[h]
\includegraphics*[width=8cm]{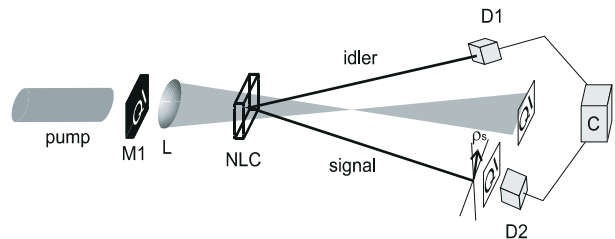}
\vspace*{0.25cm}
\caption{Experimental setup used to produce quantum images that are prepared by manipulating
the pump beam before the nonlinear crystal. The lens L is used to project the image of the mask M1 onto the detection plane.}
\label{fig1}
\end{figure}

We are going to deal with quantum images that are prepared by manipulating
the pump beam before the nonlinear crystal. In general, some transverse
pattern can be measured by displacing signal and/or idler beam detector while
performing coincidence detection. See Fig. \ref{fig1}.
Signal and idler beams freely propagate from
the crystal to the detection planes and the coincidence counting rate is given
by\cite{9}
\begin{equation}
C \left(\bbox{\rho_s},\bbox{\rho_i} \right) \propto \left|W\left[(\bbox{\rho}_i +
\bbox{\rho}_s), Z_{_{AD}}\right]\right|^2, 
\label{eq1}
\end{equation}
where $\bbox{\rho}_{s},\bbox{\rho}_{i}$ are the transverse coordinates of the
signal and idler detectors respectively, $W$ is the pump
field amplitude distribution in a plane at the same distance from the crystal as
the detectors(we have considered the same distance for both D1 and D2), 
$Z_{_{AD}}$
is the distance between crystal and detectors and the down-conversion was assumed
to be degenerated.
If one wants to send the information contained in the correlated image through
long distances, there is a drawback in using this kind of system\cite{12}, because 
signal and idler beams are naturally diverging beams as they are generated by a cavity
free finite source. The coincidence counting rate through fixed detection
surfaces is reduced, depending on the pump laser spot size inside
the crystal and on the crystal length. 
In some cases, even though the information in the correlated image is still preserved, the time for performing the measurement increases as the coincidence counting rate decreases. 
Finally, a limit is reached where the signal to noise ratio is too low and the information 
is lost.
We have calculated the coincidence counting rate, analogous to Eq.\ref{eq1},
taking into account the divergence of the signal and the idler beams:
\begin{eqnarray}
C \left(\bbox{\rho}_s,\bbox{\rho}_i \right) \propto \left(\frac{k_p}{Z_{_{AD}}}\right)^2\left|W\left[(\bbox{\rho}_i + \bbox{\rho}_s), Z_{_{AD}}\right]\right|^2,
\label{eq2}
\end{eqnarray}
where, $k_p$ is the wavenumber of the pump beam. As expected, the image is still transmitted, but now there is a factor depending 
on the inverse of the distance from the crystal to the detection plane $Z_{_{AD}}$
multiplying the coincidence counting rate. This factor is usually ignored, because
in general we are more interested in the spatial modulations of the coincidence
profile than the absolute number of coincident pairs. 
The expression above is still defined unless a multiplicative factor, the difference
from Eq.\ref{eq1} is that now the dependence on the distance $Z_{_{AD}}$
appears explicitly, showing the reduction of the number of coincident photons
per unit area.
In order to circumvent this problem, it is necessary to use collimating lenses for
signal and idler beams, but without disturbing the image entanglement. 
The combination of actions on the pump beam and on the signal and idler beams 
for obtaining
correlated images has been demonstrated in Ref. \cite{13}. It has been shown
that a di\-ffracting mask can be placed in the pump beam before the crystal,
while one imaging lens is placed  in the path of collinear twin beams for obtaining 
the correlated image. This means that the propagation of the correlated image 
was changed by the imaging lens in the signal and idler beams. The effect of this
lens in the correlated image is the same effect of a lens placed in the pump beam, 
for its intensity image.

The coincidence counting rate in this case is given by\cite{13}
\begin{eqnarray}
C\left(\bbox{\rho}_s,\bbox{\rho}_i \right) \propto
\left|W\left[\frac{O}{I}(\bbox{\rho}_i + \bbox{\rho}_s)\right]\right|^2,
\label{eq3}
\end{eqnarray}
where O is the distance between the diffracting mask and the lens
(even if part of the path is in the pump and part in the signal-idler)
and I is the distance between lens and image plane(detection plane).

\begin{figure}[h]
\includegraphics*[width=8cm]{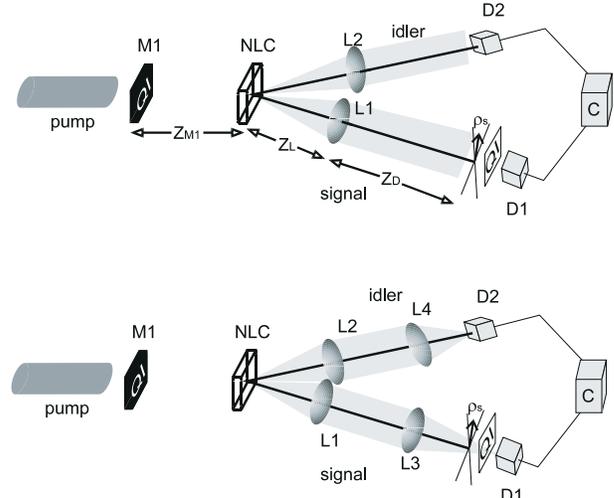}
\caption{(a) Experimental setup using collimating lenses L1 and L2 for
signal and idler beams. (b) Experimental setup with two additional lenses L3 and L4.}
\label{fig2}
\end{figure}

Eq. \ref{eq3} was obtained for the collinear and degenerate case.
However, for technical reasons, we have used non-collinear and
non-degenerate twin pairs. 
In Fig.\ref{fig2}a it is shown a typical experimental set-up, where a
diffracting mask is placed in the pump beam, before the crystal and
similar lenses are placed in the signal and idler beams. As the experimental
results are going to demonstrate, Eq. \ref{eq3} is still valid for our
set-up, because the non-degeneracy is small. The ratio between k$_i$,
k$_s$ and k$_p$/2(the degenerate wavenumber) are respectively
0.95 and 1.05 and therefore the difference is not significant. The fact
that they are non-collinear is not a problem neither, because even for the collinear
case, signal and idler are already independent modes.

The relation between variables I(image) and O(object) in Eq. \ref{eq3}
and the set-up in Fig.\ref{fig2}a is given by
O = $Z_{_{M1}}+Z_{_{L}}$ and I = $Z_{_{D}}$(The distance between lens
L2 and detector D2 in the idler beam is also $Z_{_{D}}$).
$W$ is the pump field amplitude distribution at the mask position,  $Z=0$.
Eq.\ref{eq3} shows that the information about the angular spectrum of the
pump beam scattered through mask M1 is preserved in the correlation between
signal and idler photons, after propagation through lenses L1 and L2.

If one wants to propagate the image through a long distance, the use of one
lens in the signal and one in the idler may be a complication. For long distances,
a good collimation is desired and it is necessary to place the crystal in the
focal plane of the lenses. For lenses with small focal lengths there will be
high amplification factors for the size of the image, if the detectors are very far.
One solution would be the use of very long focal length lenses, but in this case
they should be placed far from the crystal and the divergence from the crystal
to the lens would cause losses for ordinary diameters of the lenses. 
A solution for this problem is to use two or more lenses in each beam. 
For example, lenses
L1 and L2 can be designed for imaging (in coincidence) mask1 in an intermediate
plane. A second pair of lenses L3 and L4 is placed in the signal 
and idler beams. They propagate the correlated image to the detection plane.
See Fig.\ref{fig2}b.
If the distance is large, the image in the intermediate plane may have a large size,
but we must remember that signal and idler beams are still collimated, therefore
the image can be recovered and reduced by the second pair of lenses, back to
its original size. It works like a telescope.

\section{Experiment and Results}

First, we demonstrate the effect of the divergence of the
signal and idler beams in the coincidence counting rate, for a correlated
image prepared in the pump beam.
The experimental scheme is sketched in Fig.\ref{fig1}. We use a
femtosecond modelocked Ti-Safire laser, for obtaining pulses with the central
wavelength around 850nm. Second harmonic generation is performed in a
2mm long type I BBO crystal, producing pulses with wavelength centered around 425nm.

\begin{figure}[h]
\includegraphics*[width=8cm]{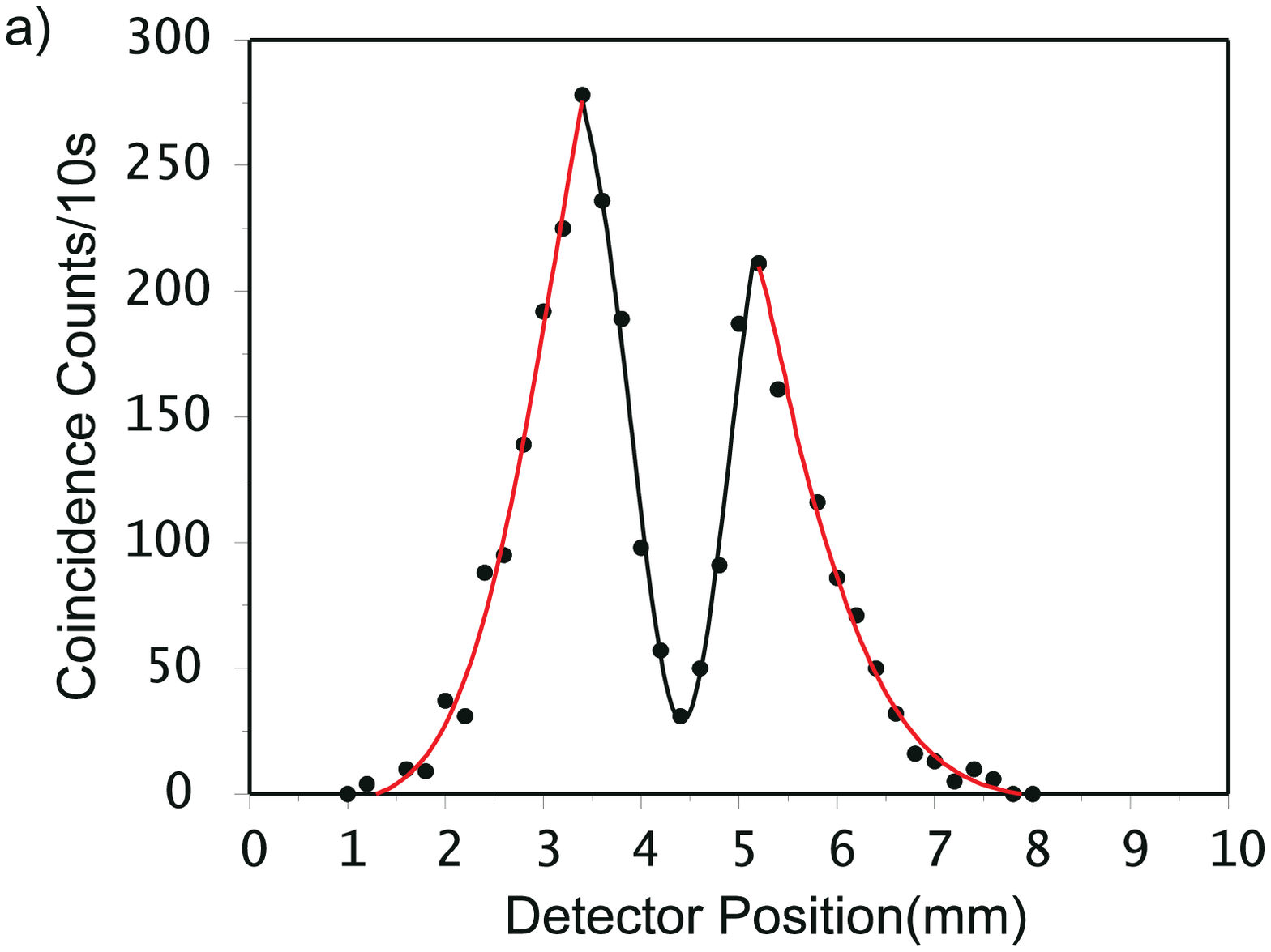}
\includegraphics*[width=8cm]{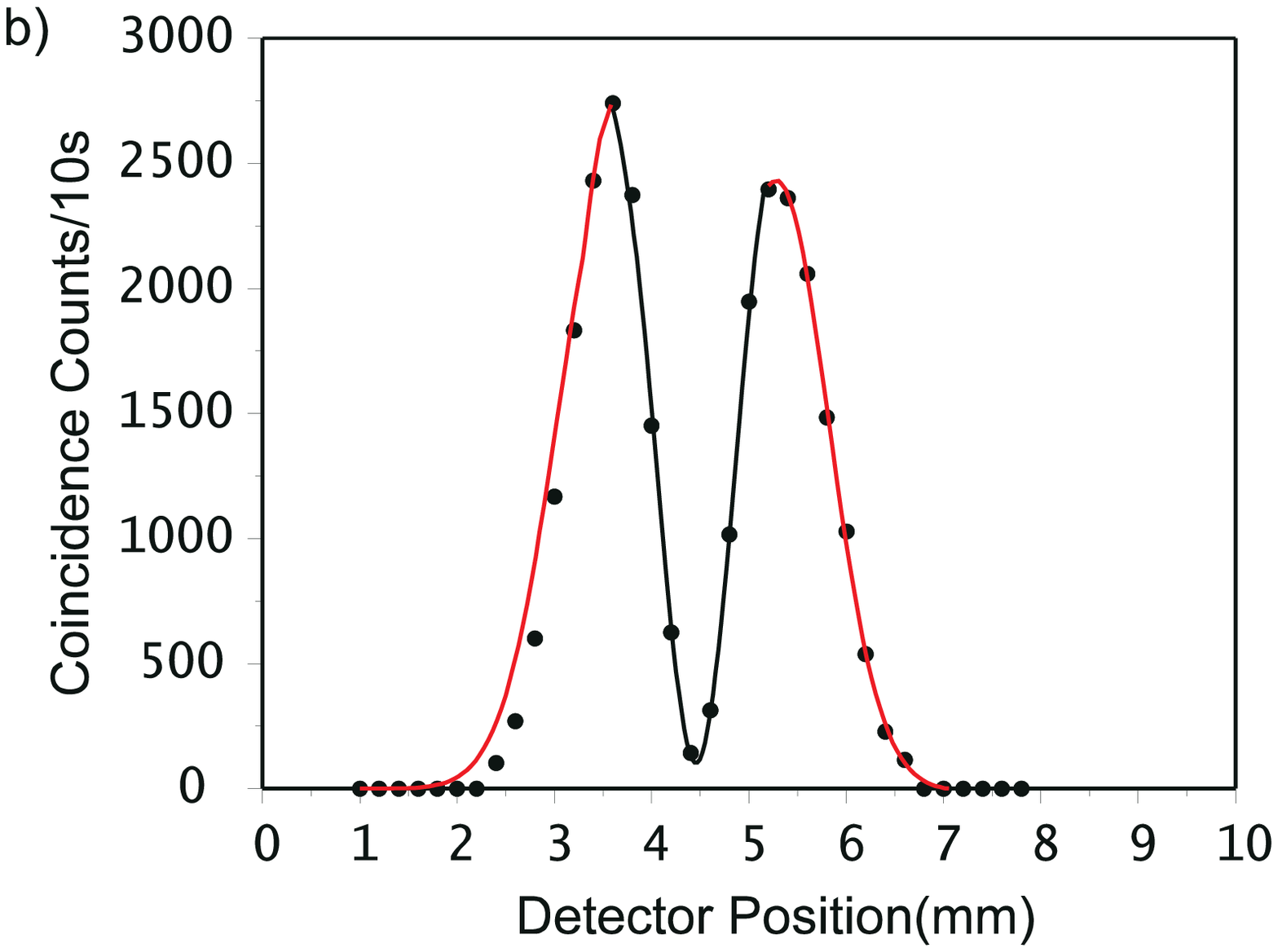}
\vspace{0.25cm}
\caption{(a) Image of the mask M1 without use of lenses in the signal and idler beams. 
(b) Image of the mask M1 with use of lenses in the signal and idler beams. Lines are only
a guide to the eye.}
\label{fig4}
\end{figure}

These pulses are used to pump a 5mm long LiIO$_{3}$ crystal, also type I
phase matched. Non collinear and non degenerate twin beams are detected
through a narrow band interference filter($\Delta \lambda \simeq 10nm$), 
centered around 890nm for the signal, and a wide band  interference filter
($\Delta \lambda \simeq 40nm$) centered around 800nm for the idler.
The image to be transmitted is obtained from the diffracting mask M1. It
consists in a thin wire. This is a very simple object and its image contains low
frequency modulations. This is convenient, as in this case it is possible to work 
with relatively large apertures in the detection, in order to have considerable coincidence 
counting rates,
and at the same time having resolution for measuring the image. If one wants
to transmit images containing small details, higher pump power and/or higher
down-conversion efficiency is required. M1 is imaged
by lens L, with the pump beam passing through the crystal, in a plane situated 
about 70cm from the crystal. The image seen in the intensity distribution of
the pump beam can be reconstructed in a correlated image by scanning signal
detector, with the idler detector fixed(or vice-versa)\cite{9}. 

This image is shown in Fig.\ref{fig4}a.
Because of the divergence of the signal and idler beams, when the distance
between the crystal and the detectors increase, the coincidence counting rate
decreases for fixed detectors surfaces. This divergence can be compensated
by placing collimating lenses L1 and L2 in the signal and idler beams, respectively,
and withdrawing lens L in the pump beam. See Fig. \ref{fig2}a. These lenses
image the quantum image in the detection plane, and at the same time they
collimate signal and idler beams. The correlated image is measured exactly in the
same way as before and the result is shown in Fig.\ref{fig4}b. Note that the
coincidence counting rate in each point is increased by a factor of about 10.

\begin{figure}[h]
\includegraphics*[width=8cm]{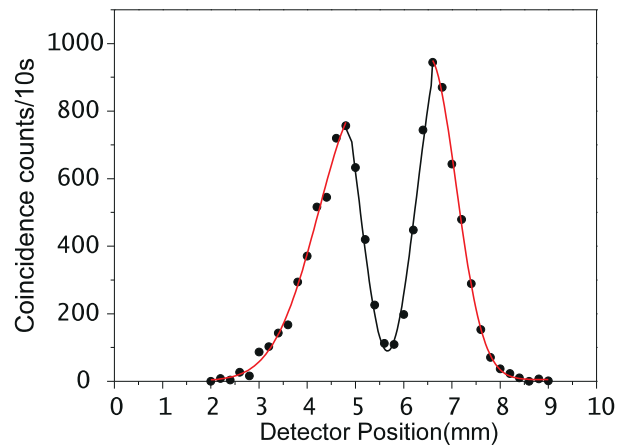}
\vspace{0.25cm}
\caption{Image of the mask M1 detected at 3m from the crystal, 
wiht use of additonal lenses L3 and L4. Lines are only a guide to the eye.}
\label{fig5}
\end{figure}

Second, we use the principle demonstrated above for detecting a correlated
image in a plane situated about 3m from the crystal. In this case, it is difficult 
to obtain the correlated image without the use of lenses in the down-converted
beams. One imaging lens placed in the pump beam, should have a focal length of
the order of meters for giving rise to an image with the same size as the object.
Therefore it should be placed far from the object and the diffraction would be
a problem for lenses with the typical diameters of about 25mm.
On the other hand, using collimating lenses in the signal and idler beams, the
quantum image was transferred with a good coincidence counting rate, comparable
to the previous measurements in Fig.\ref{fig4}b. Besides the lenses L1 and L2,
it was necessary to use other lenses, L3 and L4. See Fig.\ref{fig2}b. They
are necessary, in order to form a telescope with lenses L1 and L2 and to
keep the image with its original size. The resulting coincidence transverse
distribution is displayed in Fig.\ref{fig5}.

\section{Conclusion}

In conclusion, we have demonstrated how the natural divergence of
the signal and idler beams produced in the spontaneous parametric
down-conversion can be compensated with collimating lenses, without
loosing the quantum correlation between their angular spectra.
It is proposed that this scheme can be used for sending quantum
images through long distances. The idea of using correlated images in
cryptographic protocols have been raised in Ref.\cite{14} and that could
be associated to this long distance transmission scheme.

Financial support was provided by Brazilian agencies CNPq, PRONEX, CAPES, FAPERJ, FUJB and the Milenium Institut for Quantum Information.
\vspace*{1.5cm}

* Corresponding author.\\ E-mail address: phsr@if.ufrj.br\\


\begin{thebibliography}{99}
\bibitem{12}
Every kind of system based on twin beams from the spontaneous
parametric down-conversion is subjected to this limitation.
\end{thebibliography}
\end{document}